\documentclass{vgtc}    

\usepackage[pdf]{graphviz}
\usepackage[export]{adjustbox}
\usepackage[dvipsnames]{xcolor}

\newcommand{\hide}[1]{}

\ifpdf
  \pdfoutput=1\relax                   
  \usepackage{graphicx}                
  \DeclareGraphicsExtensions{.pdf,.png,.jpg,.jpeg} 
\else
  \usepackage{graphicx}                
  \DeclareGraphicsExtensions{.eps}     
\fi%

\graphicspath{{figures/}{pictures/}{images/}{./}} 

\usepackage{microtype}                 
\PassOptionsToPackage{warn}{textcomp}  
\usepackage{textcomp}                  
\usepackage{mathptmx}                  
\usepackage{times}                     
\usepackage{cite}                      
\usepackage{tabu}                      
\usepackage{booktabs}                  

\onlineid{0}

\vgtccategory{Research}

\vgtcinsertpkg

\title{A Comparative Analysis of Industry Human-AI Interaction Guidelines}

\author{Austin P. Wright \thanks{e-mail: apwright@gatech.edu}\\ %
\and Zijie J. Wang \thanks{e-mail: jayw@gatech.edu}\\ %
\and Haekyu Park \thanks{e-mail: haekyu@gatech.edu}\\ %
\and Grace Guo \thanks{e-mail: gguo31@gatech.edu}\\ %
\and Fabian Sperrle\thanks{e-mail: fabian.sperrle@uni-konstanz.de}\\ %
\and Mennatallah El-Assady\thanks{e-mail: mennatallah.el-assady@uni-konstanz.de}\\ %
\and Alex Endert\thanks{e-mail: endert@gatech.edu}\\ %
\and Daniel Keim \thanks{e-mail: daniel.keim@uni.kn}\\ %
\and Duen Horng (Polo) Chau\thanks{e-mail: polo@gatech.edu}} %
\affiliation{\scriptsize Georgia Institute of Technology, University of Konstanz
}

\teaser{
 \includegraphics[width=\textwidth]{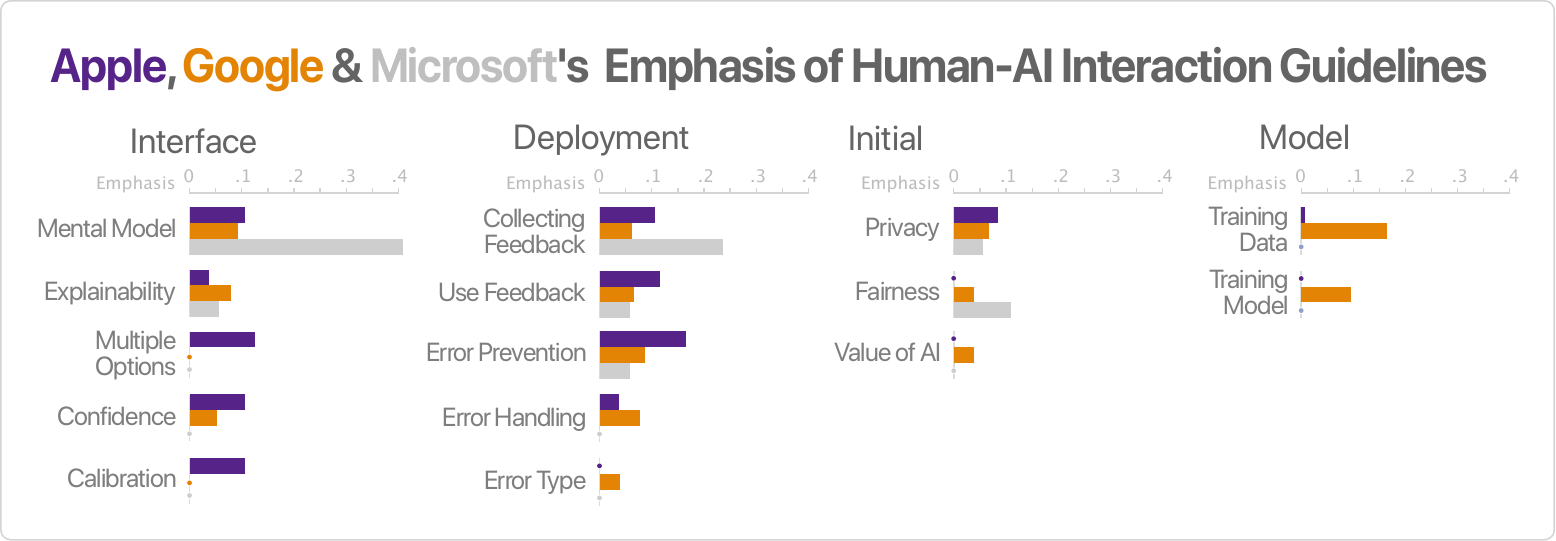}
 \caption{Relative emphasis of human-AI interaction guidelines by Apple, Google and Microsoft. A ``dot'' indicates no emphasis for a guideline subcategory.}
 \label{fig:emphasis2}
}

\abstract{
With the recent release of AI interaction guidelines from Apple, Google, and Microsoft, there is clearly interest in understanding the best practices in human-AI interaction. However, industry standards are not determined by a single company, but rather by the synthesis of knowledge from the whole community. We have surveyed all of the design guidelines from each of these major companies and developed a single, unified structure of guidelines, giving developers a centralized reference. We have then used this framework to compare each of the surveyed companies to find differences in areas of emphasis. 
Finally, we encourage people to contribute additional guidelines from other companies, academia, or individuals, to provide an open and extensible reference of AI design guidelines at \url{https://ai-open-guidelines.readthedocs.io/}. 
} 

\CCScatlist{
  \CCScatTwelve{Human-centered computing}{Human computer interaction (HCI)}{}{};
  \CCScatTwelve{Computing methodologies}{Artificial intelligence}{}{}
}

\begin{document}


\firstsection{Introduction}

\maketitle

As AI-infused systems become more common in many widely used products,
large software companies, 
including 
Apple \cite{AppleGuidelines},
Google\cite{GoogleGuidelines},
and Microsoft\cite{msftguidelines},
have recently released guidelines on designing
systems involving human-AI interaction. These guidelines from industry
sources have a great deal in common, however, they differ in key aspects
from their methodology, emphasis on different principles,  dissemination venues, and target audiences. 
No existing work has yet synthesized the information from all of these sources, leading to potentially fragmented---or even competing---standards for AI developers. 
In this work, we provide an overall analysis of all of the different
guidelines being proposed by different companies.
Our work makes the following key contributions:

\begin{enumerate}

\item 
This work provides a \textbf{comparative analysis} of design guidelines for human-AI interaction proposed by different companies. 
This includes a survey of the guidelines, comparison of each company's methodology
for developing the guidelines (in so far as it is made public), and comparison to current academic
work in human-computer interaction.
With our comparative study, AI practitioners can more easily understand both which guidelines have a consensus by the overlap and common design recommendations proposed by different companies, 
and also multi-faceted views from the differences in the companies' approaches.

\item 
Also, this work introduces a larger \textbf{inclusive taxonomy} of how all of these guidelines fit together \textbf{in a unified guideline structure} for Human-AI Interaction. This unified structure is valuable even though each of the individual sets of guidelines has its own proprietary hierarchy, as it provides an external overarching structure against which each set of guidelines can be compared, allowing us to measure the differing emphasis of each company. Furthermore, our unified structure can serve as an extensible base for future development in Human-AI Interaction.

\end{enumerate}

\noindent
We believe our work 
may serve as a helpful resource for a broad audience, including
practitioners aiming to use AI in new products, 
researchers hoping to understand Human-AI Interaction, 
and companies in understanding the current state of industry-driven best practices. This work differs from other surveys \cite{chatzimparmpas2020state,mohseni2020multidisciplinary,sperrle2020trust,hohman2018visual} on explainability and trust for AI in two ways. First, this work focuses in particular on guidelines put forth by industry; which, while they are informed by academia, do not necessarily match exactly. Furthermore, the scope of these guidelines tends to be broader beyond ensuring explainability or trust in AI but also includes guidelines for all of the other aspects that AI systems differ from other software. Therefore these guidelines can provide a more practically useful overall tool for AI developers.

\section{Survey of Guidelines}

The three most recent publications from major technology companies of guidelines have been from Google, Apple, and Microsoft. Each of these released guideline systems has a substantially different structure, emphasis, and perspective on the same central question of how to build products that use AI in a human-centered manner. This section surveys each of these industry publications and examines the context in which they were produced. 

\subsection{Microsoft}
The first set of guidelines is Microsoft's ``Guidelines for Human-AI Interaction"\cite{msftguidelines}, published at CHI 2019 in May of 2019. 
In that work, researchers from Microsoft surveyed over 168 potential guidelines originating from internal and external industry sources, public articles, and academic literature.
These guidelines were combined and re-organized into a coherent set of 18 guidelines;
all of which have a common style of ``a rule of action, containing about 3-10 words and starting with a verb"~\cite{msftguidelines}. 
Guidelines were then structured over when the guideline is relevant to the \textit{user} over the course of their interactions with the product. 

Finally, Microsoft conducted a user study with HCI practitioners to evaluate the applicability and clearness of the guidelines. This academic approach resulted in the fewest number of guidelines of the companies surveyed, however, it was the only set to outline the process explicitly on how the guidelines were developed and evaluated.

\subsection{Google}
At roughly the same time in May 2019, Google released their comprehensive set of AI interaction guidelines: Google's ``People + AI Guidebook"\cite{GoogleGuidelines}. This guidebook is based both on ``data and insights from Google product teams and academic research"\cite{GoogleGuidelines}.
While it lacks the open experimental validation of a user study, it does contain extended references to the academic literature. 
Instead of organizing the guidelines around the process for a \textit{user}, Google breaks its content down into distinct concepts that a \textit{developer} has to continuously keep in mind. These are: User Needs + Defining Success, Data Collection + Evaluation, Mental Models, Explainability + Trust, Feedback + Control, and Errors + Graceful Failure. Furthermore, the Google Guidebook takes a much longer form consisting of 113 individual guidelines, with each including more content and extensions when compared to the other publications.  
 
 \begin{figure*}
     \centering
     \includegraphics[width=\textwidth]{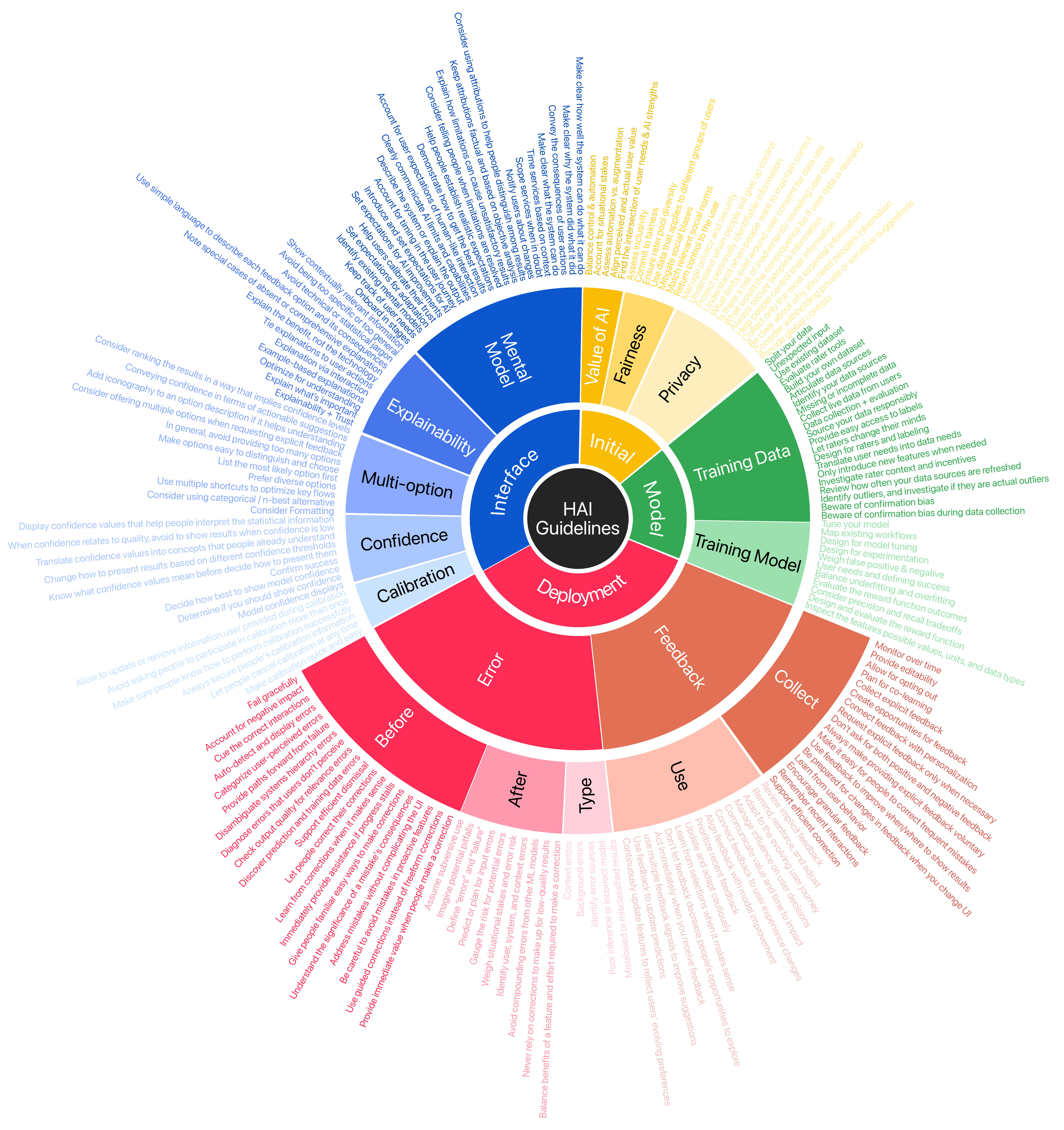}
     \caption{
        Unified Guideline Structure. 
        The inner ring consists of the higher level categorizations, 
        and further sub-categorizations developed during the affinity diagram process are shown concentrically. 
        The outermost rays consist of the specific guidelines colored based on their categorizations. Further references on each guideline and its corresponding categorization and source document can be found in the appendix.}
     \label{fig:unified_structure}
 \end{figure*}

\subsection{Apple}
Finally, in June of 2019 at WWDC'19, Apple announced its Human Interface Guidelines for Machine Learning\cite{AppleGuidelines}. These guidelines differed from the ``bottom-up" approach of the academic literature collation and user study refinement present in other guidelines. Instead, this document is a primary source of ``practitioner knowledge'', foregoing references or data, and thus is seemingly based entirely upon standing design principles within the Apple organization; which helps provide a unique and different perspective from the other more academic style works. While this style may present potential issues as a standalone document, it may help result in a greater overall synthesis of knowledge when considering all three sets of guidelines together by bringing a diversity of perspectives\cite{dur6408}. 
The document is focused on specifying how Apple's design principles are applied in the case of machine learning infused products. 
Making it comparatively more focused on aspects of user interfaces rather than AI model functionality. The 59 guidelines in the document are broken up into two main themes, the \emph{inputs} of a system and the \emph{outputs} of a system. Within each of these categories, there are further subcategories. For inputs, the guidelines focus on Explicit Feedback, Implicit Feedback, Calibration, and Corrections. Guidelines in each of these sections aim to help design the processes by which AI products ask for, collect, use, and apply user data and interactions. The sections on outputs cover Mistakes, Multiple Options, Confidence, Attribution, and Limitations. These sections all contain guidelines focused on taking the output of a model and displaying it to a user in a way that is understandable and actionable for the ultimate purpose of the product.

\begin{table}[]
    \centering
    \begin{tabular}{@{}lrrr@{}}
    \toprule
        Company & Categories & Subcategories & Guidelines\\
        \midrule
        Microsoft & 4 & N/A & 18\\
        Google & 6 & 20 & 113\\
        Apple &2 & 9& 59\\
        \bottomrule
    \end{tabular}
    \medskip
    \caption{Total number of guidelines and categories from each company surveyed.}
    \label{tab:my_label}
\end{table}
 
\section{Unified Guideline Structure}

While there are significant differences between the individual sets of guidelines, there is also substantial overlap. 
Furthermore, 
the huge amount of competing standards for desired AI systems
can wear a developer down when they try to learn and adhere to many different guidelines.
Developing a synthesis of all of the major AI guideline systems into a single comprehensive structure may make learning all of the important guidelines more straightforward, and future extensions and changes more possible. 
By fitting each company's guidelines within a larger consistent structure, the differences in emphasis between companies become readily apparent. 
This paves the way forward for the development of new guidelines and better AI-infused products.

In an affinity diagram process similar to that done by Microsoft~\cite{msftguidelines}, we separated all of the guidelines from all three sources, excluding guidelines that were not meaningful without the context of higher-level categorizations from their source document,  resulting in 194 individual guideline statements. We then conducted a card sorting exercise to find similar groups of guidelines and sort them into an affinity diagram. This resulted in twelve distinct categories of guidelines. We then repeated the process among these categories to define four high-level categories. The resulting categories are outlined below and the full hierarchy of guidelines is shown in \autoref{fig:unified_structure}.

\subsection{Initial Considerations}
These categories are generally relevant in the initial design phase of a system as things that must be considered before other development can proceed. 

\paragraph{\textit{Value of AI}}
The 5 guidelines within this category focus on understanding the value that AI may be able to bring to a product in addressing a user's need before going headfirst into development. This includes statements like ``Find the intersection of user needs and AI strengths'' and ``Balance control and automation''.  

\paragraph{\textit{Privacy / Security}}
These 13 guidelines focus on ensuring that user data is always secure and that users have the ability to control their own data. These guidelines are also broadly applicable to software more generally such as ``Always secure people’s information''  and ``Collect only the most essential information''.

\paragraph{\textit{Fairness}}
Issues of fairness in AI are an important and emerging topic in the study of human-AI interaction. However, there are comparatively few widely adopted techniques to help ensure fairness. The surveyed companies published 7 guidelines in this category, many of which are comparatively vague due to this underdeveloped area such as:  ``Mitigate social biases'' and ``Commit to fairness''. 

\subsection{Model}
These categories focus on the design of the machine learning model itself in the model design, data collection, and training procedure. 

\paragraph{\textit{Training Process}}
These 11 guidelines cover how best to handle training the model, avoiding certain kinds of errors or data issues such as ``Consider precision and recall tradeoffs'' and ``Balance underfitting and overfitting'', and ensuring the model is best optimized for the actual purpose of the product.

\paragraph{\textit{Training Data}}
While some guidelines focused on the training process, these 20 guidelines focused specifically on training datasets. These guidelines differ as questions about the data often are more high level and involve different actions, such as ``Review how often your data sources are refreshed'' or even analyzing what data is being used to ``Beware of confirmation bias''. 

\subsection{Deployment}
These categories focus on the deployment of trained models and how to handle continuous fine-tuning and errors. 

\paragraph{\textit{Errors}}
Given that AI systems fundamentally lack the kind of determinism and predictability of other software, being able to handle errors becomes especially important in this context. Therefore all three companies included extensive guidelines covering the handling of errors, resulting in a category of 26 guidelines. 
We broke down these guidelines into additional subcategories of considerations for design made before an error occurs, and systems for handling after an error occurs (``Support efficient dismissal'' and ``Learn from corrections when it makes sense''), as well as guidelines enumerating types of errors (``Mislabeled or misclassified results'' vs ``Background errors''). 

\paragraph{\textit{User Feedback / Personalization}}
There are several guidelines for how to design systems that learn specific user preferences over time, and how to give users control over this process. These are internally divided into 17 guidelines about how to collect information (``Remember recent interactions'' and ``Allow for opting out''), and 15 guidelines on using that information (``Prioritize recent feedback'' and ``Don’t let implicit feedback decrease people’s opportunities to explore'').

\begin{figure}[t]
    \centering
    \includegraphics[width=\columnwidth]{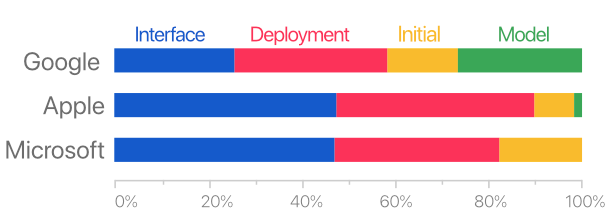}
    \caption{Relative Emphasis of Categories of Guidelines by Company}
    \label{fig:emphasis1}
\end{figure}

\subsection{Interface}
These categories focus on the design of user interactions and interfaces with AI systems.

\paragraph{\textit{Expectations / Mental Models}}
24 guidelines covered how best to design systems with user expectations in mind. 
Building intuitive mental models of how the system works is a key point of emphasis for building user trust. This includes ``Make clear why the system did what it did'' and ``Set expectations for adaptation''. 

\paragraph{\textit{Explainability}}
This category of 12 guidelines focuses on explainability. This includes trying to make the results and process of models more transparent for users. Examples include ``Show contextually relevant information'' and ``In general, avoid technical or statistical jargon''. 

\paragraph{\textit{Multiple Options}}
A key component of designing interfaces for systems with a range of potentially incorrect outputs is giving users multiple options based on model outputs. Of the 11 guidelines in this category some focus on how to display options to the user as either  ``Categorical / N-Best Alternatives'' as well as what set of options to show users who ``Prefer diverse options''. 

\paragraph{\textit{Confidence}}
A unique component of AI systems is the ability to make predictions of variable confidence. This is an important feature to expose to users the level of uncertainty in model output. The 9 guidelines covering the ways to handle uncertainty include ``avoid showing results when confidence is low'' as well as ``Decide how best to show model confidence as Categorical, N-best alternatives, or Numeric''. 

\paragraph{\textit{Calibration}}
Apple included 6 guidelines about calibrating models, which no other company talked about. Calibration differs from model training as these guidelines consider the user experience of initially fine-tuning existing models for each specific user. These include guidelines like ``Avoid asking people to participate in calibration more than once'' and ``Let people cancel calibration at any time''.

\section{Emphasis Differences}
After developing a unified structure for guidelines, we can then look back at each company's set of guidelines and compare them within this new context. To see the difference in emphasis between different companies we calculated the percentage of each company's total guidelines falling within each of our high and low-level categories. This is to control for the large difference in the number of guidelines between companies and to see how much relative emphasis each placed on different areas. 

\autoref{fig:emphasis1}  shows the distribution of high-level categories between each of the companies. This shows that the largest difference is that Google gave much more emphasis to model considerations for training data and processes, while Apple and Microsoft spent very little or no emphasis specifically on the model. Beyond that, Interface and Deployment categories dominated in roughly equal proportions at all three companies. Finally, Apple spent marginally less emphasis on initial considerations, although this effect is somewhat small. 

\autoref{fig:emphasis2} goes into more detail on which specific categories each company emphasized. The most notable is that Microsoft used nearly 40\% of its emphasis on the specific category of mental models. We can also see that Apple seems to have comparatively more emphasis in categories relating to smooth user experiences such as Error Prevention, Calibration, Confidence, and Multiple Options. 

These differences may help us understand the effects of the different methodologies used to generate these guidelines, where academic style work will tend to emphasize areas of established academic study in HCI such a mental models, while engineering-driven efforts such as Google's may focus more on the model side, and the culture and values of an organization such as Apple on seamless user experience will affect the kinds of guidelines present when developed from institutional experience. Only when surveyed together 
these differences become apparent, showing the need for comparative analysis and synthesis of all of these sources of knowledge.

\section{Discussion}

\subsection{Applications to Visualization}
Of particular note to this work are the guidelines that can be informed by visualization research. While most of the guidelines outlined fit within a much more broad HCI context, some may gain from known guidelines within visualization specifically. The interface category is the category most closely related to visualization; with subcategories such as multi-option interfaces having direct parallels in visualization concepts such as small multiples. Furthermore, data visualization research has been a hub of the best methods to achieve explainability such as the guidelines "Explanation via interaction" and "Example-based explanation" being motivating paradigms in interpretability research within visualization\cite{das2020bluff, what-if-tool}. Furthermore, there has been work within visualization assessing how best to implement many of these guidelines such as visualizing uncertainty \cite{Hullman2020} and understanding biases\cite{Wall2019,cabrera2019fairvis}. As modern AI systems are inherently data-centric, future developments in data visualization clearly inform methods of interfacing with AI products more generally.

\subsection{Limitations and Future Work}
This work has focused on published guidelines from three major companies, however, others such as IBM\cite{ibm-design-for-ai} have developed more sets of guidelines as well. Specific emphasis has been placed on fairness in particular, with guidelines coming from international organizations such as the European Union\cite{eu_ethics}, resulting in many separate specific sets of ethics guidelines that some companies are producing independently from other usability and general-purpose AI guidelines\cite{cutler2019everyday}. Further work is needed to integrate these guidelines into the structure put forth in this work. Moreover, control over these guidelines is currently being held by these few very large companies, which may have incentives to emphasize different aspects of AI than the rest of the community. Therefore these guidelines must be augmented by the community. Toward this end the guidelines developed in this work can be found at  \url{https://ai-open-guidelines.readthedocs.io/}, which puts forth an open call to collect a community-driven set of Human-Centered AI guidelines. Finally, many of these guidelines are clearly aspirational rather than practical, and thus study on the degree to which these and other companies adhere to these guidelines would be of great interest in understanding the actual effectiveness of these guidelines in the development of real products.

\section{Conclusion}

In this work, we have surveyed nearly 200 guidelines for building AI systems from three major technology companies. We have then compared them and developed a single unified taxonomy of AI guidelines. This structure allowed us to see the effects of the different approaches of these companies on what they emphasize as important. Furthermore, this structure can provide a basis of analysis for future work in developing new guidelines from industry, academia, and individuals; and synthesizing information from all of these sources to best provide a more complete reference for anyone looking to build AI systems. Finally, we have taken this work and made it open for extension, so that these guidelines are always available and determined by the community instead of solely by large companies. We hope that these guidelines then will be of use in steering the future of Human-Centered AI and assisting developers in building better AI systems.

\acknowledgments{
This work was supported in part by NSF grants IIS-1750474, IIS-1563816, CNS-1704701; gifts from Facebook, Intel, NVIDIA, Google, Symantec, Yahoo! Labs, eBay, Amazon.
This work was supported in part by Defense Advanced Research Projects Agency (DARPA). Use, duplication, or disclosure is subject to the restrictions as stated in Agreement number HR00112030001 between the Government and the Performer.
}

\bibliographystyle{abbrv-doi}

\bibliography{refs}

\newpage\clearpage

\section*{Appendix}
\subsection*{Full Set of Guidelines}

\begin{table}[h!]
\small
\begin{tabular*}{\textwidth}{@{}llll@{}}
\toprule
Category Level 1       & Category Level 2             & Company   & Guideline                                                                                                                               \\ \midrule
Initial Considerations & Value of AI                  & Google    & Find the intersection of user needs \& AI strengths.                                                                                   \\
Initial Considerations & Value of AI                  & Google    & Balance control \& automation.                                                                                                         \\
Initial Considerations & Value of AI                  & Google    & Assess automation vs. augmentation                                                                                                     \\
Initial Considerations & Value of AI                  & Google    & Align perceived and actual user value                                                                                                  \\
Initial Considerations & Value of AI                  & Google    & Account for situational stakes                                                                                                         \\
Initial Considerations & Fairness                     & Google    & Consider bias in the data collection and evaluation process                                                                            \\
Initial Considerations & Fairness                     & Google    & Assess inclusivity                                                                                                                     \\
Initial Considerations & Fairness                     & Google    & Use data that applies to different groups of users                                                                                     \\
Initial Considerations & Fairness                     & Google    & Commit to fairness                                                                                                                     \\
Initial Considerations & Fairness                     & Microsoft & Match relevant social norms.                                                                                                           \\
Initial Considerations & Fairness                     & Microsoft & Mitigate social biases.                                                                                                                \\
Initial Considerations & Fairness                     & Google    & Ensure rater pool diversity                                                                                                            \\
Initial Considerations & Privacy                      & Google    & Is there a risk of inadvertently revealing user data? What would the consequence be?                                                   \\
Initial Considerations & Privacy                      & Google    & Protect personally identifiable information                                                                                            \\
Initial Considerations & Privacy                      & Google    & Understand when people want to maintain control                                                                                        \\
Initial Considerations & Privacy                      & Google    & Understand when people will give up control                                                                                            \\
Initial Considerations & Privacy                      & Google    & What limits exist around user consent for data use                                                                                     \\
Initial Considerations & Privacy                      & Google    & Return control to the user                                                                                                             \\
Initial Considerations & Privacy                      & Google    & Manage privacy and security                                                                                                            \\
Initial Considerations & Privacy                      & Apple     & Help people control their information                                                                                                  \\
Initial Considerations & Privacy                      & Apple     & Always secure people’s information                                                                                                     \\
Initial Considerations & Privacy                      & Apple     & Collect only the most essential information.                                                                                           \\
Initial Considerations & Privacy                      & Apple     & Be clear about why you need people’s information                                                                                       \\
Initial Considerations & Privacy                      & Apple     & Consider withholding private or sensitive suggestions                                                                                  \\
Initial Considerations & Privacy                      & Microsoft & Provide global controls         \\\bottomrule                      
\end{tabular*}
\end{table}

\begin{table}[h!]
\small

\begin{tabular*}{\textwidth}{@{}llll@{}}
\toprule
Category Level 1       & Category Level 2             & Company   & Guideline                                                                                                                               \\ \midrule
Model                  & How to train your model      & Google    & Design for experimentation                                                                                                             \\
Model                  & How to train your model      & Google    & Inspect the features possible values, units, and data types                                                                            \\
Model                  & How to train your model      & Google    & Evaluate the reward function outcomes                                                                                                  \\
Model                  & How to train your model      & Google    & Weigh false positive \& negative                                                                                                       \\
Model                  & How to train your model      & Google    & Consider precision and recall tradeoffs                                                                                                \\
Model                  & How to train your model      & Google    & Balance underfitting and overfitting                                                                                                   \\
Model                  & How to train your model      & Google    & Tune your model                                                                                                                        \\
Model                  & How to train your model      & Google    & Map existing workflows                                                                                                                 \\
Model                  & How to train your model      & Google    & Design and evaluate the reward function                                                                                                \\
Model                  & How to train your model      & Google    & User needs and defining success                                                                                                        \\
Model                  & How to train your model      & Google    & Design for model tuning                                                                                                                \\
Model                  & Training Data                & Google    & Review how often your data sources are refreshed                                                                                       \\
Model                  & Training Data                & Google    & Collect live data from users                                                                                                           \\
Model                  & Training Data                & Google    & Provide easy access to labels                                                                                                          \\
Model                  & Training Data                & Google    & Use existing dataset                                                                                                                   \\
Model                  & Training Data                & Google    & Translate user needs into data needs                                                                                                   \\
Model                  & Training Data                & Google    & Only introduce new features when needed                                                                                                \\
Model                  & Training Data                & Google    & Data collection + evaluation                                                                                                           \\
Model                  & Training Data                & Google    & Identify your data sources                                                                                                             \\
Model                  & Training Data                & Google    & Identify any outliers, and investigate whether they are actual outliers or due to errors in the data                                   \\
Model                  & Training Data                & Google    & Source your data responsibly                                                                                                           \\
Model                  & Training Data                & Google    & Build your own dataset                                                                                                                 \\
Model                  & Training Data                & Google    & Design for raters and labeling                                                                                                         \\
Model                  & Training Data                & Google    & Split your data                                                                                                                        \\
Model                  & Training Data                & Google    & Let raters change their minds                                                                                                          \\
Model                  & Training Data                & Google    & Evaluate rater tools                                                                                                                   \\
Model                  & Training Data                & Google    & Missing or incomplete data                                                                                                             \\
Model                  & Training Data                & Google    & Unexpected input                                                                                                                       \\
Model                  & Training Data                & Google    & Investigate rater context and incentives                                                                                               \\
Model                  & Training Data                & Google    & Articulate data sources                                                                                                                \\
Model                  & Training Data                & Apple     & Beware of confirmation bias                                                                                                            \\\bottomrule
\end{tabular*}
\end{table}

\newpage
\clearpage
\newpage

\begin{table}[h!]
\small
\begin{tabular*}{\textwidth}{@{}lllp{10cm}@{}}
\toprule
Category Level 1       & Category Level 2             & Company   & Guideline                                                                                                                               \\ \midrule
Interface              & Explainability               & Google    & Explain the benefit, not the technology                                                                                                \\
Interface              & Explainability               & Google    & Use simple, direct language to describe each explicit feedback option and its consequences                                             \\
Interface              & Explainability               & Google    & Optimize for understanding                                                                                                             \\
Interface              & Explainability               & Google    & Explainability + Trust                                                                                                                 \\
Interface              & Explainability               & Google    & Note special cases of absent or comprehensive explanation                                                                              \\
Interface              & Explainability               & Google    & Explanation via interaction                                                                                                            \\
Interface              & Explainability               & Google    & Example-based explanations                                                                                                             \\
Interface              & Explainability               & Google    & Explain what’s important                                                                                                               \\
Interface              & Explainability               & Google    & Tie explanations to user actions                                                                                                       \\
Interface              & Explainability               & Apple     & In general, avoid technical or statistical jargon                                                                                      \\
Interface              & Explainability               & Apple     & Avoid being too specific or too general                                                                                                \\
Interface              & Explainability               & Microsoft & Show contextually relevant information                                                                                                 \\
Interface              & Confidence                   & Google    & Model confidence displays                                                                                                              \\
Interface              & Confidence                   & Google    & Decide how best to show model confidence                                                                                               \\
Interface              & Confidence                   & Google    & Categorical                                                                                                                            \\
Interface              & Confidence                   & Google    & N-best alternatives                                                                                                                    \\
Interface              & Confidence                   & Google    & Numeric                                                                                                                                \\
Interface              & Confidence                   & Google    & Determine if you should show confidence                                                                                                \\
Interface              & Confidence                   & Apple     & When you know that confidence values correspond to result quality, you generally want to avoid showing results when confidence is low. \\
Interface              & Confidence                   & Apple     & Consider changing how you present results based on different confidence thresholds                                                     \\
Interface              & Confidence                   & Apple     & In general, translate confidence values into concepts that people already understand.                                                  \\
Interface              & Confidence                   & Apple     & Know what your confidence values mean before you decide how to present them                                                            \\
Interface              & Confidence                   & Apple     & In scenarios where people expect statistical or numerical information, display confidence values that help them interpret the results. \\
Interface              & Confidence                   & Apple     & Confirm success                                                                                                                        \\
Interface              & Expectations / Mental Models & Google    & Onboard in stages.                                                                                                                     \\
Interface              & Expectations / Mental Models & Google    & Help users calibrate their trust.                                                                                                      \\
Interface              & Expectations / Mental Models & Google    & Introduce and set expectations for AI                                                                                                  \\
Interface              & Expectations / Mental Models & Google    & Set expectations for AI improvements                                                                                                   \\
Interface              & Expectations / Mental Models & Google    & Account for timing in the user journey                                                                                                 \\
Interface              & Expectations / Mental Models & Google    & Keep track of user needs                                                                                                               \\
Interface              & Expectations / Mental Models & Google    & Identify existing mental models                                                                                                        \\
Interface              & Expectations / Mental Models & Google    & Clearly communicate AI limits and capabilities                                                                                         \\
Interface              & Expectations / Mental Models & Google    & Set expectations for adaptation.                                                                                                       \\
Interface              & Expectations / Mental Models & Google    & Describe the system or explain the output                                                                                              \\
Interface              & Expectations / Mental Models & Google    & Account for user expectations of human-like interaction.                                                                               \\
Interface              & Expectations / Mental Models & Apple     & Consider using attributions to help people distinguish among results.                                                                  \\
Interface              & Expectations / Mental Models & Apple     & Keep attributions factual and based on objective analysis.                                                                             \\
Interface              & Expectations / Mental Models & Apple     & Help people establish realistic expectations.                                                                                          \\
Interface              & Expectations / Mental Models & Apple     & Explain how limitations can cause unsatisfactory results                                                                               \\
Interface              & Expectations / Mental Models & Apple     & Consider telling people when limitations are resolved                                                                                  \\
Interface              & Expectations / Mental Models & Apple     & Demonstrate how to get the best results                                                                                                \\
Interface              & Expectations / Mental Models & Microsoft & Make clear what the system can do.                                                                                                     \\
Interface              & Expectations / Mental Models & Microsoft & Make clear how well the system can do what it can do                                                                                   \\
Interface              & Expectations / Mental Models & Microsoft & Make clear why the system did what it did.                                                                                             \\
Interface              & Expectations / Mental Models & Microsoft & Convey the consequences of user actions.                                                                                               \\
Interface              & Expectations / Mental Models & Microsoft & Notify users about changes.                                                                                                            \\
Interface              & Expectations / Mental Models & Microsoft & Scope services when in doubt.                                                                                                          \\
Interface              & Expectations / Mental Models & Microsoft & Time services based on context.                                                                                                        \\
Interface              & Calibration                  & Apple     & Avoid asking people to participate in calibration more than once.                                                                      \\
Interface              & Calibration                  & Apple     & Make calibration quick and easy                                                                                                        \\
Interface              & Calibration                  & Apple     & Make sure people know how to perform calibration successfully.                                                                         \\
Interface              & Calibration                  & Apple     & Let people cancel calibration at any time.                                                                                             \\
Interface              & Calibration                  & Apple     & Give people a way to update or remove information they provided during calibration.                                                    \\
Interface              & Calibration                  & Apple     & Always secure people's calibration information                                                                                         \\
Interface              & Multiple Options             & Google    & Categorical / N-Best Alternatives                                                                                                      \\
Interface              & Multiple Options             & Google    & Consider Formatting                                                                                                                    \\
Interface              & Multiple Options             & Google    & Use multiple shortcuts to optimize key flows                                                                                           \\
Interface              & Multiple Options             & Apple     & Whenever possible, help people make decisions by conveying confidence in terms of actionable suggestions.                              \\
Interface              & Multiple Options             & Apple     & List the most likely option first.                                                                                                     \\
Interface              & Multiple Options             & Apple     & In situations where attributions aren't helpful, consider ranking or ordering the results in a way that implies confidence levels      \\
Interface              & Multiple Options             & Apple     & Consider offering multiple options when requesting explicit feedback.                                                                  \\
Interface              & Multiple Options             & Apple     & In general, avoid providing too many options                                                                                           \\
Interface              & Multiple Options             & Apple     & Prefer diverse options                                                                                                                 \\
Interface              & Multiple Options             & Apple     & Make options easy to distinguish and choose                                                                                            \\
Interface              & Multiple Options             & Apple     & Add iconography to an option description if it helps people understand it.                                                             \\ \bottomrule
\end{tabular*}
\end{table}

\newpage
\clearpage
\newpage

\begin{table}[p]
\small

\begin{tabular*}{\textwidth}{@{}lllp{10cm}@{}}
\toprule
Category Level 1       & Category Level 2             & Company   & Guideline                                                                                                                               \\ \midrule
Deployment             & Error Prevention             & Google    & Account for negative impact                                                                                                            \\
Deployment             & Error Prevention             & Google    & Auto-detect and display errors                                                                                                         \\
Deployment             & Error Prevention             & Google    & Disambiguate systems hierarchy errors                                                                                                  \\
Deployment             & Error Prevention             & Google    & Diagnose errors that users don’t perceive                                                                                              \\
Deployment             & Error Prevention             & Google    & Check output quality for relevance errors                                                                                              \\
Deployment             & Error Prevention             & Google    & Fail gracefully                                                                                                                        \\
Deployment             & Error Prevention             & Google    & Discover prediction and training data errors                                                                                           \\
Deployment             & Error Prevention             & Google    & Cue the correct interactions                                                                                                           \\
Deployment             & Error Prevention             & Google    & Categorize user-perceived errors                                                                                                       \\
Deployment             & Error Prevention             & Google    & Provide paths forward from failure                                                                                                     \\
Deployment             & Error Prevention             & Apple     & Understand the significance of a mistake’s consequences                                                                                \\
Deployment             & Error Prevention             & Apple     & As you work on reducing mistakes in one area, always consider the effect your work has on other areas and overall accuracy             \\
Deployment             & Error Prevention             & Apple     & When possible, address mistakes without complicating the UI                                                                            \\
Deployment             & Error Prevention             & Apple     & Learn from corrections when it makes sense                                                                                             \\
Deployment             & Error Prevention             & Apple     & When possible, use guided corrections instead of freeform corrections                                                                  \\
Deployment             & Error Prevention             & Apple     & Let people correct their corrections                                                                                                   \\
Deployment             & Error Prevention             & Apple     & Provide immediate value when people make a correction                                                                                  \\
Deployment             & Error Prevention             & Apple     & Give people familiar easy ways to make corrections                                                                                     \\
Deployment             & Error Prevention             & Apple     & Immediately provide assistance if progress stalls                                                                                      \\
Deployment             & Error Prevention             & Apple     & Be especially careful to avoid mistakes in proactive features                                                                          \\
Deployment             & Error Prevention             & Microsoft & Support efficient dismissal                                                                                                            \\
Deployment             & Error types                  & Google    & Identify error sources.                                                                                                                \\
Deployment             & Error types                  & Google    & Background errors.                                                                                                                     \\
Deployment             & Error types                  & Google    & Context errors                                                                                                                         \\
Deployment             & Error types                  & Google    & Mislabeled or misclassified results                                                                                                    \\
Deployment             & Error types                  & Google    & Poor inference or incorrect model                                                                                                      \\
Deployment             & Error Handling               & Google    & Assume subversive use                                                                                                                  \\
Deployment             & Error Handling               & Google    & Imagine potential pitfalls                                                                                                             \\
Deployment             & Error Handling               & Google    & Gauge the risk for potential errors                                                                                                    \\
Deployment             & Error Handling               & Google    & Identify user, system, and context errors                                                                                              \\
Deployment             & Error Handling               & Google    & Weigh situational stakes and error risk                                                                                                \\
Deployment             & Error Handling               & Google    & Avoid compounding errors from other ML models                                                                                          \\
Deployment             & Error Handling               & Google    & Define “errors” and “failure”                                                                                                          \\
Deployment             & Error Handling               & Google    & Predict or plan for input errors                                                                                                       \\
Deployment             & Error Handling               & Apple     & Never rely on corrections to make up for low-quality results                                                                           \\
Deployment             & Error Handling               & Apple     & Always balance the benefits of a feature with the effort required to make a correction                                                 \\
Deployment             & Collecting Feedback          & Google    & Collect explicit feedback.                                                                                                             \\
Deployment             & Collecting Feedback          & Google    & Monitor over time.                                                                                                                     \\
Deployment             & Collecting Feedback          & Google    & Allow for opting out.                                                                                                                  \\
Deployment             & Collecting Feedback          & Google    & Plan for co-learning.                                                                                                                  \\
Deployment             & Collecting Feedback          & Google    & Connect feedback with personalization.                                                                                                 \\
Deployment             & Collecting Feedback          & Google    & Create opportunities for feedback.                                                                                                     \\
Deployment             & Collecting Feedback          & Google    & Provide editability.                                                                                                                   \\
Deployment             & Collecting Feedback          & Apple     & Be prepared for changes in implicit feedback when you make changes to your app’s UI.                                                   \\
Deployment             & Collecting Feedback          & Apple     & Don’t ask for both positive and negative feedback.                                                                                     \\
Deployment             & Collecting Feedback          & Apple     & Make it easy for people to correct frequent or predictable mistakes.                                                                   \\
Deployment             & Collecting Feedback          & Apple     & Always make providing explicit feedback a voluntary task.                                                                              \\
Deployment             & Collecting Feedback          & Apple     & Request explicit feedback only when necessary.                                                                                         \\
Deployment             & Collecting Feedback          & Apple     & Consider using explicit feedback to help improve when and where you show results.                                                      \\
Deployment             & Collecting Feedback          & Microsoft & Remember recent interactions.                                                                                                          \\
Deployment             & Collecting Feedback          & Microsoft & Encourage granular feedback.                                                                                                           \\
Deployment             & Collecting Feedback          & Microsoft & Support efficient correction.                                                                                                          \\
Deployment             & Collecting Feedback          & Microsoft & Learn from user behavior.                                                                                                              \\
Deployment             & Addressing / using Feedback  & Google    & Review implicit feedback.                                                                                                              \\
Deployment             & Addressing / using Feedback  & Google    & Adapt to the evolving user journey.                                                                                                    \\
Deployment             & Addressing / using Feedback  & Google    & Remind, reinforce, and adjust.                                                                                                         \\
Deployment             & Addressing / using Feedback  & Google    & Communicate value and time to impact.                                                                                                  \\
Deployment             & Addressing / using Feedback  & Google    & Align feedback with model improvement.                                                                                                 \\
Deployment             & Addressing / using Feedback  & Google    & Manage influence on user decisions.                                                                                                    \\
Deployment             & Addressing / using Feedback  & Google    & Connect feedback to user experience changes.                                                                                           \\
Deployment             & Addressing / using Feedback  & Apple     & When possible, use multiple feedback signals to improve suggestions and mitigatie mistakes.                                            \\
Deployment             & Addressing / using Feedback  & Apple     & Prioritize recent feedback.                                                                                                            \\
Deployment             & Addressing / using Feedback  & Apple     & Learn from selections when it makes sense.                                                                                             \\
Deployment             & Addressing / using Feedback  & Apple     & Update and adapt cautiously.                                                                                                           \\
Deployment             & Addressing / using Feedback  & Apple     & Use feedback to update predictions on a cadence that matches the user's mental model of the feature.                                   \\
Deployment             & Addressing / using Feedback  & Apple     & Act immediately when you receive explicit feedback and persist the resulting changes.                                                  \\
Deployment             & Addressing / using Feedback  & Apple     & Don’t let implicit feedback decrease people’s opportunities to explore.                                                                \\
Deployment             & Addressing / using Feedback  & Microsoft & Continuously update your feature to reflect people’s evolving interests and preferences.                                               \\ \bottomrule
\end{tabular*}
 \end{table}

\end{document}